\documentclass[a4paper]{article}

\usepackage{INTERSPEECH2022}
\usepackage{float}
\usepackage{multirow}
\usepackage{amsfonts,amssymb}
\usepackage{hyperref}
\usepackage{enumitem}
\usepackage{indentfirst}
\usepackage{makecell}

\title{AdaSpeech 4: Adaptive Text to Speech in Zero-Shot Scenarios}
\name{Yihan Wu$^{1}$, Xu Tan$^{2,*}$, Bohan Li$^{3}$, Lei He$^{3}$, Sheng Zhao$^{3}$, Ruihua Song$^{1,*}$\thanks{$^*$ Work done during the first author is interning at Microsoft Research Asia. Corresponding author: songruihua\_bloon@outlook.com, xuta@microsoft.com.}, Tao Qin$^{2}$, Tie-Yan Liu$^{2}$}
\address{$^1$ Gaoling School of Artificial Intelligence, Renmin University of China\\
$^2$ Microsoft Research Asia,\ 
$^3$ Microsoft Azure Speech}
\email{yihanwu@ruc.edu.cn, \{xuta, bohli, helei, taoqin, szhao, tyliu\}@microsoft.com}

\begin{document}

\maketitle
\begin{abstract}
Adaptive text to speech (TTS) can synthesize new voices in zero-shot scenarios efficiently, by using a well-trained source TTS model without adapting it on the speech data of new speakers. 
Considering seen and unseen speakers have diverse characteristics, zero-shot adaptive TTS requires strong generalization ability on speaker characteristics, which brings modeling challenges.  
In this paper, we develop AdaSpeech 4~\footnote{AdaSpeech series aim for adaptive TTS, where AdaSpeech is a basic model backbone for efficient TTS adaptation, AdaSpeech 2 is for TTS adaptation with untranscribed data, and AdaSpeech 3 is for TTS adaptation in spontaneous style. We develop AdaSpeech 4 based on the basic model backbone of AdaSpeech.}, a zero-shot adaptive TTS system for high-quality speech synthesis. We model the speaker characteristics systematically to improve the generalization on new speakers. Generally, the modeling of speaker characteristics can be categorized into three steps: extracting speaker representation, taking this speaker representation as condition, and synthesizing speech$/$mel-spectrogram given this speaker representation. Accordingly, we improve the modeling in three steps: 
1) To extract speaker representation with better generalization, we factorize the speaker characteristics into basis vectors and extract speaker representation by weighted combining of these basis vectors through attention. 2) We leverage conditional layer normalization to integrate the extracted speaker representation to TTS model. 
3) We propose a novel supervision loss based on the distribution of basis vectors to maintain the corresponding speaker characteristics in generated mel-spectrograms. Without any fine-tuning, AdaSpeech 4 achieves better voice quality and similarity than baselines in multiple datasets.


\end{abstract}

\section{Introduction}
\label{sec:intro}
Neural text to speech (TTS)~\cite{tan2021survey} models can synthesize high quality human voice when being trained with a large amount of single-speaker or multi-speaker datasets~\cite{skerry2018towards,shen2018natural,ren2019fastspeech,ren2020fastspeech2,liu2021delightfultts,gibiansky2017deepvoice2,ping2017deepvoice3,chen2020multispeech}. 
When synthesizing speech for new speakers, few-shot adaptive TTS~\cite{chen2021adaspeech,yan2021ada2,yan2021ada3,huang2021metatts} is usually adopted by first training a source TTS model on a large multi-speaker dataset and then fine-tuning this model on a few speech data of target speakers. Although few-shot adaptive TTS achieves good similarity and voice quality on target speakers, it has two limitations: 1) it needs some training data of target speakers, which are hard to obtain from consumers; 2) it needs fine-tuning a source TTS model on target data, which incurs much computation cost when serving a lot of new speakers. Such a situation is common for commercialized TTS services (e.g., Microsoft Azure, Google Cloud, Amazon Web Services, etc.). 

Zero-shot adaptive TTS~\cite{min2021meta,wu2022msdtron} can generate a new voice by only modeling the speaker characteristics from a reference speech, without adapting the source TTS model on the speech data of new speakers. Although zero-shot adaptive TTS is both data and computation efficient, it faces big challenges for achieving good voice quality. Specifically, considering target/unseen speakers and source/seen speakers can have many diverse characteristics, it requires the source TTS model to have strong generalization ability on speaker characteristics. In this paper, we first categorize the modeling of speaker characteristics in current TTS systems into several steps, and then propose a new system to improve the generalization ability on speaker characteristics at each step and thus achieve better zero-shot quality. 

Basically, the modeling of speaker characteristics in TTS can be categorized into three steps: 1) extracting speaker representation from target speaker; 2) taking the extracted representation as a condition to TTS model; and 3) generating target mel-spectrogram given this speaker representation. For the first step, previous zero-shot TTS models~\cite{casanova2021yourtts,cooper2020zeroshotemb,chien2021investigating} usually leverage a speaker/reference encoder to extract speaker representation of a target speaker. Most efforts focus on improving the speaker encoder to extract a more adaptable speaker representation. However, speaker representation is hard to be precisely extracted in zero-shot scenarios since various factors such as timbre, speaking style, and prosody need to be considered. For the second step, previous models usually concatenate or add the extracted speaker embedding with the hidden output of phoneme encoder and then take them as the input of the decoder, which causes a mismatched decoder input when the speaker embedding is not precisely extracted in zero-shot scenarios, and affects the generalization ability. In the third step, most previous works have no explicit guarantee that the generated mel-spectrograms follow the same speaker characteristics as the extracted speaker representation, where the situation is even worse in zero-shot scenarios. Some previous works~\cite{xin2021disentangled,wang2020bilevel} apply speaker classification loss on generated mel-spectrograms as a supervision to ensure the synthesized speech to be more similar to reference speech, which, however, does not bring much improvement in speaker similarity~\cite{casanova2021yourtts}.


\begin{figure*}[t]
\setlength{\abovecaptionskip}{0.1cm}
\setlength{\belowcaptionskip}{0.05cm}
  \centering
  \includegraphics[width=0.80\linewidth]{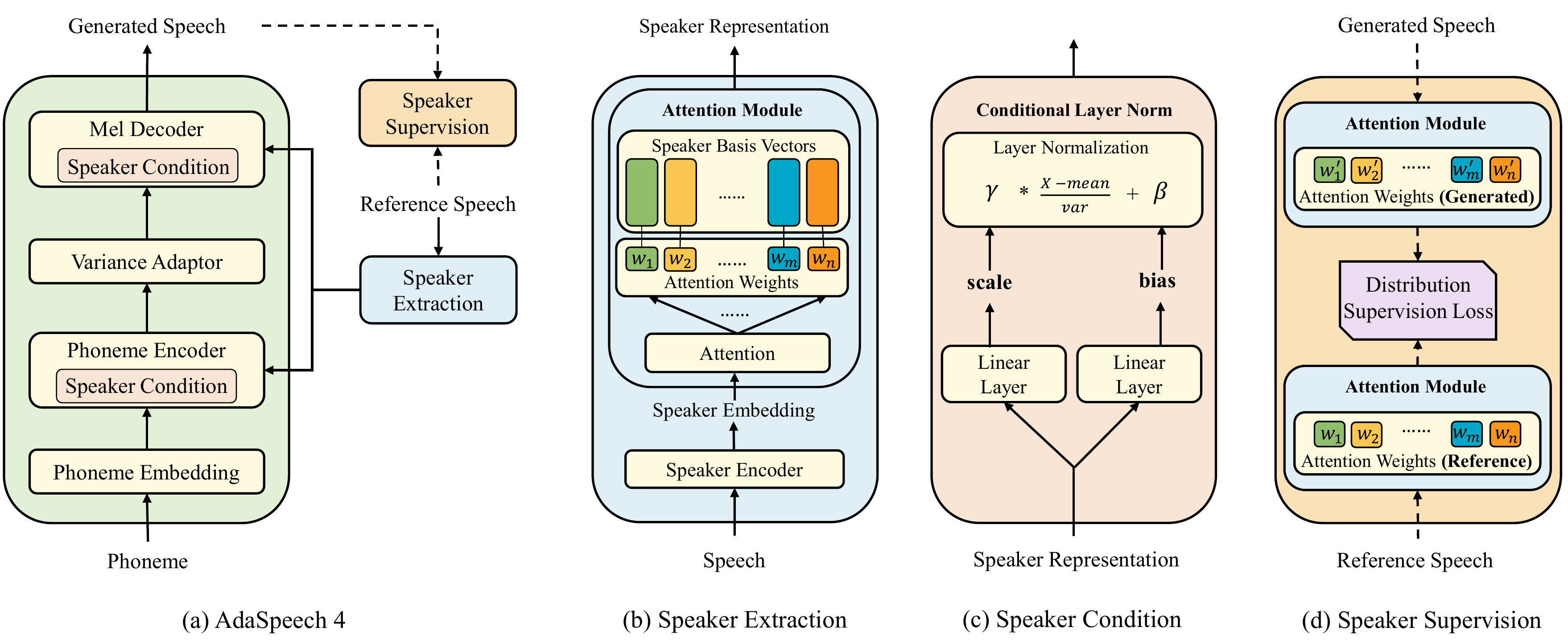}
  \caption{The architecture of AdaSpeech 4, where the basic model backbone is based on AdaSpeech~\cite{chen2021adaspeech}, but with three systematic designs to improve the generation ability on speaker characteristics in zero-shot scenarios. Note that we also use phoneme-level acoustic condition modeling as in AdaSpeech, which is not shown for simplicity. Dashed lines in Figure (a) and (d) represent speaker supervision is only used in the training stage, and ``speech'' refers to mel-spectrograms here. }
  \label{fig:fig1}
  \vspace{-1.5em}
\end{figure*}

Based on the above analyses, in this paper, we develop AdaSpeech 4, an adaptive TTS model for high-quality speech synthesis in zero-shot scenarios. Based on the model structure of AdaSpeech~\cite{chen2021adaspeech}, we improve the generalization ability on new speakers in three steps:
\begin{itemize}[leftmargin=*]
\item To extract speaker representation with better generalization, we factorize the speaker characteristics into basis vectors and extract speaker representation by weighted combining of these basis vectors through attention, which can ensure good generalization on new speakers in zero-shot scenarios. To ensure the basic vectors to be dissimilar (ideally need to be orthogonal), we initialize basis vectors with the cluster centers (by $k$-means~\cite{macqueen1967kmeans}) of pre-trained speaker embeddings and encourage the basis vectors to be dissimilar with a regularization loss.  

\item We employ conditional layer normalization to take the extracted speaker representation as input to the phoneme encoder and mel-spectrogram decoder of TTS model. In conditional layer normalization, the scale and bias vectors are generated by model parameters conditioned on extracted speaker representation, which improves the generalization on new speakers compared with directly taking extracted speaker representation as decoder input. 

\item We propose a novel supervision loss based on the distribution of basis vectors to ensure the generated mel-spectrograms to be similar to the reference speech in terms of speaker representation, which helps to generate more similar mel-spectrograms in zero-shot scenarios.


\end{itemize}

We train AdaSpeech 4 on LibriTTS datasets and conduct zero-shot synthesis on new speakers in LibriTTS, VCTK, and LJSpeech datasets. Experiment results show that AdaSpeech 4 achieves better voice quality and voice similarity in terms of MOS (Mean Opinion Score) and SMOS (Similarity Mean Opinion Score) than baseline methods in zero-shot scenarios. Ablation studies and method analyses verify the effectiveness of each design in AdaSpeech 4. Audio samples are available at \url{https://speechresearch.github.io/adaspeech4/}. 
\begin{table*}[t]
\footnotesize
  \caption{The MOS and SMOS scores with 95\% confidence on LibriTTS, VCTK, and LJSpeech.}
  \label{tab:overal_mos_smos}
  \centering
  \begin{tabular}{l|ccc|ccc}
    \toprule
    \textbf{Metric} &  & \textbf{SMOS$ \ (\uparrow )$} &  &    & \textbf{MOS$ \ (\uparrow )$} & \\ 
    \midrule
    \textbf{Dataset} & \textbf{LibriTTS}  & \textbf{VCTK} & \textbf{LJSpeech} & \textbf{LibriTTS} & \textbf{VCTK} & \textbf{LJSpeech}   \\
    \midrule
    GT & $ 4.09 \pm 0.11 $  & $ 4.17 \pm 0.10 $ & $4.08 \pm 0.10 $ & $3.45 \pm 0.12$  & $3.69 \pm 0.13$ & $3.67 \pm 0.13$\\
    GT\ mel + Vocoder & $3.97 \pm 0.10$  & $4.13 \pm 0.09$ & $4.03 \pm 0.09$ & $ 3.42 \pm 0.12 $ & $ 3.68 \pm 0.14 $ & $3.62 \pm 0.12 $  \\
    \midrule
    FastSpeech 2 (vanilla)~\cite{ren2020fastspeech2} & $3.33 \pm 0.13 $  & $ 3.23 \pm 0.12 $& $3.26 \pm 0.13 $  & $3.12 \pm 0.14 $ & $ 3.54 \pm 0.14 $ & $3.16 \pm 0.12 $   \\
   FastSpeech 2 (d-vector)~\cite{ren2020fastspeech2} & $ 3.12 \pm 0.14 $ & $ 2.98 \pm 0.12 $ & $ 2.80 \pm 0.12 $  & $3.08 \pm 0.14 $ & $2.63 \pm 0.13 $ & $ 3.53 \pm 0.14 $ \\
    StyleSpeech~\cite{min2021meta} & $3.64 \pm 0.13 $  & $3.81 \pm 0.11  $ & $3.58 \pm 0.13 $ & $3.24 \pm 0.13 $  & $ 3.58 \pm 0.14  $ & $3.30 \pm 0.18 $   \\
    AdaSpeech (zero-shot) ~\cite{chen2021adaspeech}  & $ 3.62 \pm 0.14$ & $ 3.79 \pm 0.10 $ & $3.54 \pm 0.12$ & $ 3.22 \pm 0.13$ & $ 3.63 \pm 0.14 $ & $3.35 \pm 0.17$  \\
    \midrule
    AdaSpeech 4 & \bm{$3.88 \pm 0.11 $}  & \bm{$3.86 \pm 0.10 $}  & \bm{ $ 3.68 \pm 0.10$ }  &\bm{ $3.34 \pm 0.12$} & \bm{$ 3.66 \pm 0.13 $} & \bm{$ 3.37 \pm 0.11 $} \\
    \bottomrule
  \end{tabular}
  \vspace{-1.3em}
\end{table*}

\section{Proposed Method}
The whole architecture of AdaSpeech 4 is shown in Figure~\ref{fig:fig1}(a), where the model backbone is based on AdaSpeech~\cite{chen2021adaspeech}, a non-autoregressive TTS model with specifically designed acoustic condition modeling for few-shot adaptation. Based on the categorization of the three key steps in speaker characteristics modeling as in Section~\ref{sec:intro}, we further improve the model's generalization ability to new speakers in zero-shot scenarios correspondingly.

First, we employ a set of basis vectors to represent speaker characteristics and extract more generalized speaker representation through attention, as shown in Figure~\ref{fig:fig1}(b). Second, we integrate the extracted speaker representation to TTS model by conditional layer normalization to minimize the generalization difficulty for unseen speakers, as shown in Figure~\ref{fig:fig1}(c). Third, we leverage a supervision loss based on the distribution of the above basis vectors to improve the controllability of speaker characteristics in zero-shot scenarios, as shown in Figure~\ref{fig:fig1}(d).

\subsection{Extracting Speaker Representation by Basis Vectors}
Speaker representation is hard to extract precisely in zero-shot scenarios due to complicated speaker characteristics need to be captured. A better way is to represent speaker characteristics with a set of basis vectors and extract speaker representation through the weighted combination of these basis vectors. Since these basic vectors are learnt from all speakers during training, they should have enough representation capability for different characteristics. When extracting speaker representation in zero-shot scenarios, we do not need to generate a representation from scratch, but just need to generate new combination weights for these basis vector, which has more generalization capability to unseen speakers. 
Inspired by ~\cite{wang2018style} which models speech style by global style tokens for expressive speech synthesis, we leverage a similar pipeline to learn basis vectors and extract speaker representations (as shown in Figure~\ref{fig:fig1}(b)): we use the speaker embedding $S$ generated by speaker encoder as query, and attend to the basis vector through Q-K-V attention~\cite{Vaswani2017attention} to extract the speaker representation $E$~\cite{wu2022msdtron,zheng2021zeroshotGST}. 
\begin{equation}
\setlength\abovedisplayskip{3pt}
\setlength\belowdisplayskip{3pt}
     E = Attention (SW^{Q}, BW^{K}, BW^{V}),
     \label{eq:attention1}
\end{equation}
\begin{equation}
\setlength\abovedisplayskip{2pt}
\setlength\belowdisplayskip{2pt}
     Attention (Q, K, V) = softmax (\frac{Q K^{T}}{\sqrt{d}}) V,
\label{eq:attention}
\end{equation}
where $W^{Q}$,$W^{K}$,$W^{V}$ are all trainable matrices, $B$ denotes basis vectors, $Q$, $K$, $V$ in Equation~\ref{eq:attention} are attention queries, keys, and values, and $d$ is the dimension of $S$.

To ensure the representation capability on seen speakers and generalization capability on unseen speakers, the basis vectors should be dissimilar to spread out to the whole space of speaker characteristics. To this end, we propose two improvements: 1) We initialize the basis vectors with the cluster centers of speaker embeddings extracted by the speaker encoder. 
2) We leverage a regularization loss to prevent each basis vector to be similar to each other.
Specifically, we extract speaker embeddings from all speech training data by the speaker encoder, and partition them into $N$ clusters by $k$-means clustering. Here, $N$ is a hyper-parameter and is equal to the number of basis vectors. Then we initialize basis vectors with the cluster centers to make the vectors different from each other and thus ensure the representation capability from the beginning of training. Furthermore, we employ a regularization loss to minimize the similarity among basis vectors, which is computed as,
\begin{equation}
\setlength\abovedisplayskip{3pt}
\setlength\belowdisplayskip{3pt}
    \mathcal{L}_{reg}=\frac{1}{N (N-1)}\sum_{i}^{N} \sum_{j, j \neq i}^{N} cos<b_{i}, b_{j}>
    \label{eq:cos_loss}
\end{equation}
where $b_i,\ b_j \in B$ refers to the $i$-$th$ and $j$-$th$ basis vector respectively. By encouraging the vectors to be dissimilar in the training process, they can spread out to the whole space of speaker characteristics to improve the generalization capability on unseen speakers in zero-shot scenarios.

\subsection{Integrating Speaker Representation with Conditional Layer Normalization}
\label{sec:speaker condition}
Since speaker representation is hard to be estimated precisely for unseen speakers, taking inaccurate speaker representation as the input of decoder will cause a mismatch between source model training and zero-shot synthesis. Therefore, we explore a better condition method to take speaker representation as model input to improve the generalization of TTS model in zero-shot scenarios (as shown in Figure~\ref{fig:fig1}(c)). Specifically, we determine the scale and bias vectors in layer normalization with the extracted speaker representation using a small conditional network~\cite{min2021meta}, i.e., a linear layer $W^{\gamma}$ for scale and $W^{\beta}$ for bias. Both linear layers take extracted speaker representation $E$ as input and output adaptive scale and bias vectors as follows:
\begin{equation}
\setlength\abovedisplayskip{3pt}
\setlength\belowdisplayskip{3pt}
    \gamma=E \times W^{\gamma}, \quad \beta=E \times W^{\beta}.
\end{equation}
We substitute the conventional layer normalizations in each self-attention and feed-forward network in Transformer with our conditional layer normalization.

Different from AdaSpeech~\cite{chen2021adaspeech}, for zero-shot scenarios, we employ speaker representation extracted from reference speech as the input of conditional layer normalization instead of speaker embedding projected from speaker ID. Besides, as the output of phoneme encoder is used to predict variance information (e.g., pitch, duration) related to speaker identity through variance adaptor, it is also required to have strong generalization ability on speaker characteristics. Therefore, instead of only employing conditional layer normalization in mel decoder in AdaSpeech~\cite{chen2021adaspeech}, we leverage conditional layer normalization both in the phoneme encoder and mel decoder, which shows better zero-shot effectiveness in experiments.

\subsection{Supervising Speaker Representation in Synthesized Speech with Distribution Loss}
\label{sec:speaker supervision}
There is no explicit guarantee that the generated mel-spectrograms follow the same speaker characteristics as the extracted speaker representation, which could affect the similarity of the generated speech, especially in zero-shot scenarios. Thus, we propose a novel supervision loss based on the distribution of basis vectors to  minimize the difference of speaker characteristics between reference mel-spectrograms and synthesized mel-spectrograms (as shown in Figure~\ref{fig:fig1}(d)). Specifically, we employ KL divergence loss to minimize the distance between the attention weights of the reference speech (calculated by the attention module in Equation~\ref{eq:attention}) and that of the generated speech:
\begin{equation}
\setlength\abovedisplayskip{3pt}
\setlength\belowdisplayskip{3pt}
    \mathcal{L}_{dist} = \sum_{i=1}^{N} w_{i} log\frac{w_{i}}{w_{i}^{'}},
    \label{eq:kl}
\end{equation}
where $w_{i} = softmax (\frac{S W^{Q} (b_{i} W^{K})^{T}}{\sqrt{d}})$ according to Equation~\ref{eq:attention1} and~\ref{eq:attention}, which denotes the attention weights of speaker embedding $S$ from reference speech to the basis vector $b_{i}$. $w_{i}^{'}$ follows the same calculation method but from generated speech instead.

Unlike previous work which maximizes the similarity score based on speech embedding, our proposed distribution-level loss requires the reference speaker representation and the generated speaker representation to have the similar distribution over the shared basis vectors. It maintains speaker representation in synthesized speech in a more controllable way, which benefits zero-shot scenarios. 

\section{Experiments and Results}

\subsection{Datasets and Experiments Settings}

\noindent \textbf{Datasets.} 
We train AdaSpeech 4 on LibriTTS dataset~\cite{zen2019libritts}, which is a multi-speaker TTS corpus derived from LibriSpeech~\cite{panayotov2015librispeech}. We split the dataset into training, validation, and test sets. All speakers in the test set are unseen during training. To evaluate generalization abilities in various acoustic conditions, we also conduct zero-shot synthesis in VCTK~\cite{veaux2016vctk} (a multi-speaker dataset) and LJSpeech~\cite{Ito2017LJ} (a single-speaker dataset). We randomly sample ten speakers (including five men and five women) from multi-speaker datasets (i.e., LibriTTS and VCTK) and the one speaker from single-speaker dataset (i.e., LJSpeech). Then we randomly select one audio from each speaker as reference and synthesize 15 sentences for human evaluation. The way of preprocessing on the speech and text data follows AdaSpeech~\cite{chen2021adaspeech}.


\noindent \textbf{Model configurations.}
The model configurations of AdaSpeech 4 follow AdaSpeech~\cite{chen2021adaspeech} unless otherwise stated. 
The speaker encoder~\cite{skerry2018towards,wang2018style} is a 6-layer convolution network, where each layer is composed of 3 × 3 filters with 2 × 2 strides, using ``same'' padding and ReLU activations. Batch normalization~\cite{ioffe2015batchnorm} is applied to every layer. Output channels for 6 convolutional layers are 32, 32, 64, 64, 128, 128. The number of basis vectors $N$ is set to 2000 in our experiments.

\noindent \textbf{Training and inference pipeline.}
The training process is divided into three stages, and mel-spectrogram reconstruction loss is used in all three stages.
1) We pre-train a multi-speaker TTS model with the speaker encoder. Then we employ this pre-trained speaker encoder to extract the speaker embedding of all the training utterances and use their $k$-means centers ($N=2000$ clusters) as the initialization of basis vectors.
2) We continue to train our AdaSpeech 4 with all the model parameters optimized with additional regularization loss (Equation~\ref{eq:cos_loss}). 
3) To maintain the speaker representation in generated mel-spectrograms, we fix the speaker encoder and optimize the rest parameters with the distribution loss (Equation~\ref{eq:kl}).  
\begin{table}
\footnotesize
  \caption{The SMOS and CMOS scores with 95\% confidence on LibriTTS for ablation study.}
  \label{tab:ablation_smos_cmos}
  \centering
  \setlength{\tabcolsep}{0.5mm}{
  \begin{tabular}{l|l|l|c|c}
    \toprule
    \textbf{\thead{Ablation\\ Modules}}& \textbf{ID} & \makecell[c]{\textbf{Settings}} & \textbf{SMOS$ \ (\uparrow )$}  & \textbf{CMOS$ \ (\uparrow )$} \\ 
    \midrule
    & \#1 & AdaSpeech 4 & \bm{$3.88 \pm 0.11$} & \bm{$0$} \\
    \midrule
    \multirow{3}{*}{\thead{Speaker \\ Extraction}} & \#2 & \#1 $-$ $k$-means init &$3.84 \pm 0.11$ &$-\ 0.36$ \\
   & \#3 & \#1 $-$ $\mathcal{L}_{reg}$ & $3.67 \pm 0.11$ & $-\ 0.46$\\
   & \#4 & \#1 $-$ basis vectors & $3.62 \pm 0.12$ & $-\ 0.17$\\
     \midrule
    \multirow{2}{*}{\thead{Speaker\\ Condition}} & \#5 & \#1 $-$ encoder CLN &$3.72 \pm 0.12 $ & $-\ 0.21$\\
    & \#6 & \#5 $-$ decoder CLN &$3.70 \pm 0.11 $ & $-\ 0.36$ \\
     \midrule
   \multirow{3}{*}{\thead{Speaker\\ Supervision}} & \#7 &  \#1 $-$ $\mathcal{L}_{dist}$ &$3.64 \pm 0.12 $ & $-\ 0.06$ \\
    & \#8 &  \#7 $+$ $\mathcal{L}_{cos}$ & $3.69 \pm 0.13 $ & $-\ 0.08$\\
    & \#9 &  \#1 $+$ $\mathcal{L}_{cos}$ & $3.71 \pm 0.12 $ & $-\ 0.06$ \\
    \bottomrule
  \end{tabular}}
  \vspace{-1.2em}
\end{table}

\subsection{Similarity and Quality Comparison with Baselines}
To evaluate the voice quality and similarity, we conduct subjective listening tests including MOS (Mean Opinion Score) and SMOS (Similarity MOS) on Microsoft crowd-sourcing platform. Each sentence is listened by 20 native judgers. For VCTK and LibriTTS, we average the MOS and SMOS scores of multiple speakers as the final scores.
We compare the synthesized speech of AdaSpeech 4 with several baselines. 1) GT: the ground-truth recordings; 2) GT Mel + Vocoder: using ground-truth mel-spectrogram to synthesize waveform with HiFi-GAN vocoder~\cite{kong2020hifigan}; 3) StyleSpeech~\cite{min2021meta}; 4) FastSpeech 2 (vanilla): a FastSpeech 2 based multi-speaker TTS model which adds the output of speaker encoder as speaker embedding to the phoneme encoder's output; 5) FastSpeech 2 (d-vector): similar to 4), except for 
leveraging d-vector as speaker embedding; 6) AdaSpeech (zero-shot): the implementation of AdaSpeech in zero-shot scenarios, i.e., without fine-tuning. All the above baselines use HiFi-GAN as the vocoder to generate waveforms.
The MOS and SMOS results are shown in Table~\ref{tab:overal_mos_smos} respectively.
We observe that AdaSpeech 4 achieves good improvements in SMOS in both three datasets while maintaining good or achieving slightly better voice quality in terms of MOS. 

\subsection{Ablation Studies}
\begin{figure}[t]
  \centering
  \includegraphics[width=\linewidth]{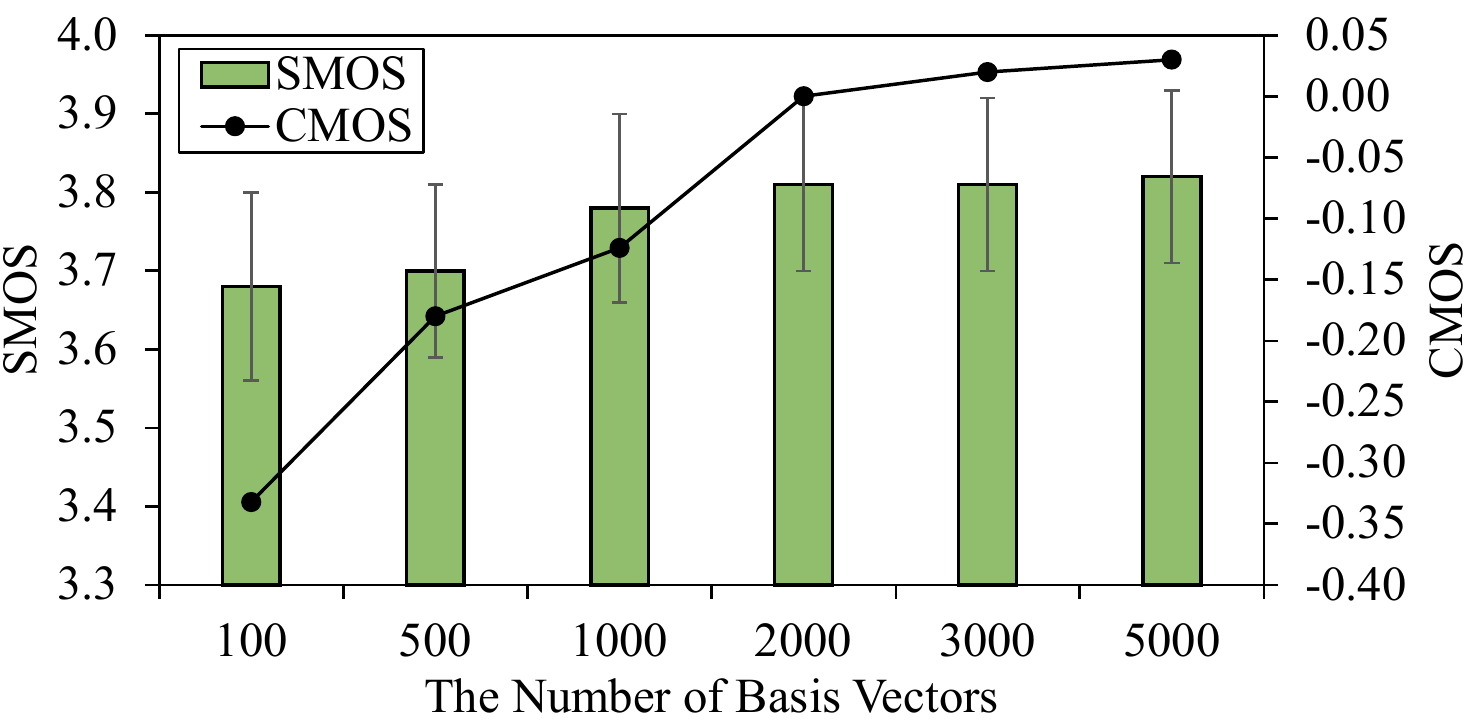}
  \caption{The SMOS and CMOS of different number of basis vectors on LibriTTS test set.}
  \label{fig:basis_num}
  \vspace{-1.2em}
\end{figure}

In this section, we conduct ablation studies to verify the effectiveness of each component in AdaSpeech 4. As shown in Table~\ref{tab:ablation_smos_cmos}, we can have following observations:
\begin{itemize}[leftmargin=*]
\item
\textbf{Speaker extraction.} When removing the initialization operation with $k$-means clustering centers (\#2), it leads to a voice quality drop with $-$ 0.36 CMOS. Both removing regularization loss (\#3) and speaker basis vectors (\#4) result in SMOS and CMOS drop, demonstrating the effectiveness of each component in the speaker extraction module. Removing regularization loss (\#3) brings the largest drop, i.e., $-$ 0.46 in terms of CMOS.
\item
\textbf{Speaker condition.} Discarding conditional layer normalization (CLN) in phoneme encoder (\#5) or both in mel decoder and phoneme encoder (\#6) impairs voice quality and speaker similarity, which verifies the effectiveness of conditional layer normalization in AdaSpeech 4.
\item
\textbf{Speaker supervision.} Discarding distribution loss (\#7) leads to a voice similarity drop in terms of SMOS. Besides, we apply a cosine similarity loss (denoted as $\mathcal{L}_{cos}$) between reference speaker embedding and generated speaker embedding as supervision loss, which is applied in many previous works~\cite{casanova2021yourtts,xin21crossada}. However, employing this embedding-level supervision loss 
alone (\#8) or jointly (\#9) does not bring obvious gain.

\item
\textbf{The number of basis vectors.} As the number of basis vectors determines the variety of speaker characteristics, we further investigate the zero-shot quality under different number of basis vectors on LibriTTS test set. As shown in Figure~\ref{fig:basis_num}, the voice quality and similarity continuously drops when the number of basis vectors decreases from 2000, while there is no obvious gain when the number of basis vectors is greater than 2000. Thus, we choose 2000 in our experiments.

\end{itemize}

\section{Conclusion}
In this paper, we develop AdaSpeech 4, an adaptive TTS system for high-quality speech synthesis in zero-shot scenarios. We categorize the modeling of speaker characteristics into three steps and improve its generalization ability in a systematic way. Specifically, we extract speaker representation by basis vectors, integrate the extracted speaker
representation to TTS model by conditional layer normalization, and maintain speaker representation in synthesized speech with a novel distribution-level supervision loss. Experiment results demonstrate that AdaSpeech 4 can synthesize speech with high quality and similarity in zero-shot scenarios. For future work, we will evaluate AdaSpeech 4 in more diverse speaker characteristics and explore advanced techniques to improve the prosody and expressiveness of synthesized speech in zero-shot scenarios.

\vfill\pagebreak

\bibliographystyle{IEEEtran}

\bibliography{mybib}

\end{document}